\documentclass[10pt,letterpaper]{article}

\usepackage[margin=0.85in]{geometry}
\usepackage[T1]{fontenc}
\usepackage{lmodern}
\usepackage{amsmath,amssymb,amsthm}
\usepackage{graphicx}
\usepackage{booktabs}
\usepackage{natbib}
\usepackage{xcolor}
\definecolor{linknavy}{RGB}{0,51,102}
\usepackage[colorlinks=true,allcolors=linknavy]{hyperref}
\IfFileExists{microtype.sty}{\usepackage{microtype}}{}

\newcommand{\E}{\mathbb{E}}
\newcommand{\Prob}{\mathbb{P}}
\DeclareMathOperator{\Var}{Var}
\newcommand{\one}{\mathbf{1}}

\makeatletter
\renewcommand\section{\@startsection{section}{1}{\z@}%
  {-2.0ex \@plus -0.5ex \@minus -.2ex}{0.8ex \@plus .2ex}%
  {\normalfont\large\bfseries}}
\renewcommand\subsection{\@startsection{subsection}{2}{\z@}%
  {-1.6ex \@plus -0.4ex \@minus -.2ex}{0.6ex \@plus .2ex}%
  {\normalfont\normalsize\bfseries}}
\renewcommand\paragraph{\@startsection{paragraph}{4}{\z@}%
  {1.4ex \@plus .4ex \@minus .2ex}{-0.6em}{\normalfont\normalsize\bfseries}}
\makeatother
\title{\vspace{-1.5em}On the Benefit of Blocking for Online Experiments with Skewed Data}
\author{Abbas Zaidi, Matthew Reimherr, Rina Friedberg, Fabian Martinez and Richard Mudd
        \\[0.3em] Meta Platforms Inc}
\date{}

\begin{document}

\maketitle

\section{Introduction}

Online randomized controlled experiments (``A/B tests'') empower decision-making across the information technology (IT) sector, supporting both deployment decisions, such as a new user experience, and evaluations of policy choices, such as invoice billing \citep{xu2015infrastructure}. Design and analysis choices vary significantly across experiments; some, for instance, contend with highly skewed pre-experiment characteristics. Despite these variations, product decision processes depend on a company's ability to perform high-quality inference across its collection of A/B tests \citep{kohavi2020trustworthy}. The quality of A/B testing as a platform for decision making depends on summary statistics that represent operating characteristics, such as Type I and Type II error rates, that enable ideation, iteration, and launch. This quality can be severely hampered when \emph{covariate} imbalances arise, a particular commonality when their distribution is highly skewed. The problem is typically handled in the IT sector through analysis-side changes such as regression adjustment, though these come with their own challenges, including small-sample considerations and asymptotic approximations that are not readily satisfied.

We advocate for designing a balanced experiment upfront by way of \textit{blocking}, a mechanism that guarantees baseline protection against severe sampling imbalances. It removes the need to rely on potentially flawed post-hoc statistical models to salvage the data after the fact. Outside of IT, across agriculture \citep{box1980ra}, education policy \citep{krueger1999experimental}, and modern multi-center clinical trials \citep{zheng2008multi}, blocking has consistently proven its benefit by neutralizing environmental and operational confounders before an experiment even begins. Our work contributes to the increasingly resource-constrained field of online experimentation, where prioritizing structural prevention over post-hoc correction has become essential for maintaining system efficiency. Specifically, when massive sample sizes are not feasible, such as when studying a highly skewed population of small businesses, blocking provides a path to statistical efficiency without relying on oft-fragile, post-hoc assumptions.

The existing evidence on how much improvement stratification results in is bracketed by two very different answers. Empirically, \citet{gualavisi2025stratification} study several large-scale experiments stratified into 48 to 64 cells and find standard errors improving by only five to ten percent, with pre- and post-stratification performing almost identically; they conclude that moderate numbers of strata capture most of the available benefit. Theoretically, \citet{bai2026efficiency} show that stratifying into blocks of fixed size makes the unadjusted estimator asymptotically efficient, so that the gain arrives by design rather than through adjustment. What separates these two answers is a question of rate: how much of the available gain is realized as a function of the number of blocks, and therefore how fine a design must be before blocking becomes beneficial. That is the question we take up, in the regime where the outcome is heavily skewed, which the empirical evidence above does not cover.

\section{Technical Details}
\label{sec:setup}

\subsection{Notation and Estimand}

Consider a single experiment. Units are indexed by $i = 1, \dots, N$ and are drawn independently from a population. Each unit carries a \textit{block} label $G_i \in \{1, \dots, J\}$, a deterministic function of pre-experiment characteristics; a treatment assignment $Z_i \in \{0, 1\}$; and a pair of potential outcomes $(Y_i(1), Y_i(0))$, of which we observe $Y_i = Z_i Y_i(1) + (1 - Z_i) Y_i(0)$ \citep{ding2018causal}. We write $n_j = \#\{i : G_i = j\}$ for the number of units in block $j$ and $n_{aj} = \#\{i : G_i = j,\, Z_i = a\}$ for the number assigned to arm $a$ within it. Under \textit{blocked assignment} the arm counts $n_{1j}$ are fixed in advance and randomization is carried out independently within each block, so that
\begin{equation}
\bigl(Y_i(1), Y_i(0)\bigr) \perp Z_i \;\bigm|\; G_i.
\label{eq:ignorability}
\end{equation}
Conditioning on the block label is therefore not a matter of notational convenience: equation~\eqref{eq:ignorability} is the level at which the design delivers \emph{ignorability}, and so every quantity below is defined within a block and aggregated afterwards. Accordingly, we define the block-conditional treatment effect $\theta_j$ and the block share $\omega_j$ as $\theta_j = \E[\, Y_i(1) - Y_i(0) \mid G_i = j \,]$ and $\omega_j = \Prob(G_i = j)$, so that, by the law of total expectation, our target of inference is
\begin{equation}
\theta = \E\bigl[Y_i(1) - Y_i(0)\bigr] = \sum_{j=1}^{J} \omega_j \, \theta_j.
\label{eq:tower}
\end{equation}
Equation~\eqref{eq:tower} is an identity rather than a modelling assumption; it holds for any blocking variable, and the weights are fixed by the design rather than estimated. What is at stake here is characterized in Section~\ref{sec:buys}.

\subsection{Estimation and Inference}

\paragraph{Assumptions and Block-Varying Variance.} Alongside the blocked assignment of \eqref{eq:ignorability} we assume that there is no interference between units, so that $Y_i$ responds only to $Z_i$; that every block receives both arms, $0 < n_{1j} < n_j$ for all $j$, which the design enforces but which becomes harder to satisfy as $J$ grows; and that the within-block potential-outcome variances $\sigma^2_{aj} = \Var\bigl(Y_i(a) \mid G_i = j\bigr)$ are finite. We do \textit{not} assume that these are common across blocks. Blocking variables in this setting are typically measures of pre-experiment scale: a seller's prior sales volume, a content creator's subscriber base. Variables of this kind are themselves heavily skewed, so the blocks they induce cut the population up very unevenly. When the outcome is measured on the same scale as the blocking variable, which is the usual case, the pattern is ordered: dispersion grows with the block index, and $\sigma^2_{aJ}$ can exceed $\sigma^2_{a1}$ by orders of magnitude. We therefore carry the index $j$ on every second moment below.

\paragraph{Estimator.} Within block $j$, \eqref{eq:ignorability} makes the difference in means unbiased for $\theta_j$, and \eqref{eq:tower} prescribes how to combine:
\begin{equation}
\hat\theta_j = \frac{1}{n_{1j}} \sum_{i \,:\, G_i = j,\, Z_i = 1} \! Y_i
\;-\; \frac{1}{n_{0j}} \sum_{i \,:\, G_i = j,\, Z_i = 0} \! Y_i ,
\qquad
\hat\theta = \sum_{j=1}^{J} \hat\omega_j \hat\theta_j ,
\quad \hat\omega_j = \frac{n_j}{N}.
\label{eq:estimator}
\end{equation}
Because the weights are fixed by the design rather than estimated from the outcome, $\hat\theta$ is unbiased for $\theta$ irrespective of $\sigma^2_{aj}$: block-varying variance results in lost precision, not validity. The block label is in practice a discretization of an underlying pre-experiment covariate $x_i$, retaining the variation between cells and discarding the variation within them. Adjusting for $x_i$ in the block recovers some of it:
\begin{equation}
Y_i = \sum_{j=1}^{J} \one\{G_i = j\}\Bigl[\, \beta_j + \gamma_j Z_i
+ \delta_j (x_i - \bar x_j) + \eta_j Z_i (x_i - \bar x_j) \,\Bigr] + \varepsilon_i .
\label{eq:regression}
\end{equation}
This is the within-block form of the interacted adjustment of \citet{lin2013agnostic}; dropping $\delta_j$ and $\eta_j$ recovers \eqref{eq:estimator}. The reason to raise adjustment at all is that it and the design are competing for the same variation, which makes them substitutes: the finer the blocks, the less within-block variation in $x_i$ is left for the adjustment to use \citep{bai2026efficiency}.

\paragraph{Inference.} Assignment is independent across blocks, so cross-block covariances vanish and the variance is a weighted sum of block-terms, each on its own scale:
\begin{equation}
\widehat{\Var}(\hat\theta) = \sum_{j=1}^{J} \hat\omega_j^{2}
\biggl( \frac{s_{1j}^{2}}{n_{1j}} + \frac{s_{0j}^{2}}{n_{0j}} \biggr),
\qquad
\mathrm{CI}_{1-q} = \hat\theta \pm z_{1-q/2}\, \widehat{\Var}(\hat\theta)^{1/2},
\label{eq:variance}
\end{equation}
where $s^2_{aj}$ is the within-block, within-arm sample variance. Under equal allocation $n_{1j} = n_{0j} = n_j/2$ and a common within-block scale $\sigma^2_j$, \eqref{eq:variance} collapses to $\Var(\hat\theta) = (4/N)\sum_j \omega_j \sigma^2_j = (4/N)\,\E[\Var(Y \mid G)]$. This is where block-varying variance bites hardest: a block holding only a small share of the population can still dominate the sum when its variance is large enough. Two caveats attach to \eqref{eq:variance}. The normal approximation's accuracy turns on the shape of the within-block distribution, so blocks remaining strongly asymmetric have poorer coverage. More consequentially, \eqref{eq:variance} is valid for the block structure used in the \textit{analysis}: if assignment blocks more finely than the analysis partition, it overstates true sampling variability.

\section{Mathematical Intuition}
\label{sec:buys}

Blocking is usually justified by precision alone. It does at least two further things that are not precision arguments at all, which we isolate below. Throughout, allocation is equal and $\sigma^2_j$ is common to both arms.

\paragraph{Precision: the gain is the variance share of the label.} Under complete randomization the difference in means has variance $(4/N)\Var(Y)$, against $(4/N)\,\E[\Var(Y \mid G)]$ for the blocked estimator. The law of total variance gives:
\begin{equation}
\frac{\Var(\hat\theta_{\mathrm{block}})}{\Var(\hat\theta_{\mathrm{complete}})}
\;=\; \frac{\E[\Var(Y \mid G)]}{\Var(Y)} \;=\; 1 - R^2_G ,
\qquad
R^2_G \;:=\; \frac{\Var(\E[Y \mid G])}{\Var(Y)} .
\label{eq:ratio}
\end{equation}
Applying the same decomposition twice separates the covariate $x$ from the discrete labels $G$:
\begin{equation}
R^2_G \;=\; R^2_x \cdot \kappa_J ,
\qquad
R^2_x = \frac{\Var\bigl(\E[Y \mid x]\bigr)}{\Var(Y)} ,
\qquad
\kappa_J = \frac{\Var\bigl(\E[Y \mid G]\bigr)}{\Var\bigl(\E[Y \mid x]\bigr)} .
\label{eq:decomp}
\end{equation}
The first factor is how much of the outcome the covariate explains; the second is how much of that a $J$-cell partition retains. Skew acts directly on the second factor: equal-count cells are wide where the population is sparse, bounding the potential gain from blocking. 

\paragraph{Estimand: only the design weights estimate $\theta$.} The quantity a business needs is $\theta$, the average effect over the population. By \eqref{eq:tower} that is $\sum_j \omega_j \theta_j$. Any other weighting estimates a different quantity. Writing $v_j = 1/\sigma^2_j$, the natural efficient alternative $w_j \propto \omega_j v_j$ targets $\theta_{\mathrm{iv}} = \sum_j \omega_j v_j \theta_j \big/ \sum_j \omega_j v_j$, and
\begin{equation}
\theta_{\mathrm{iv}} - \theta
\;=\; \frac{\sum_{j} \omega_j v_j \,(\theta_j - \theta)}{\sum_{j} \omega_j v_j} .
\label{eq:ivbias}
\end{equation}
Under skew, $\sigma^2_j$ increases with the block index, so $v_j$ falls, discounting exactly the blocks holding the largest units. Where those blocks carry larger effects, $\theta_{\mathrm{iv}}$ sits below $\theta$. There is no bias-variance trade available here because the requirement is an interval covering $\theta$. Efficiency weighting provides tight precision about the wrong quantity. Fixing weights by design gives up the efficiency available only when $\theta_j \equiv \theta$, and buys the estimand in every other case.

\paragraph{Blocking localizes operations that would otherwise be global.} Skewed outcomes are commonly winsorized. Writing $W_c(y) = \min(y, c)$ and $(u)_+ = \max(u, 0)$, a global clip at threshold $c$ distorts the estimand by $\theta - \theta^{\mathrm{glob}}(c) = \E[(Y(1) - c)_+] - \E[(Y(0) - c)_+]$. For a proportional lift $\lambda$, the fraction of the effect destroyed is the fraction of total outcome volume lying above the threshold:
\begin{equation}
\frac{\theta - \theta^{\mathrm{glob}}(c)}{\theta}
\;\approx\; \frac{\E\bigl[Y \one\{Y > c\}\bigr]}{\E[Y]} .
\label{eq:volshare}
\end{equation}
Volume above a high quantile is exactly what a heavy tail has in abundance. The same calculation run inside a block gives the local rule. Let $c_j$ be the $(1-q)$ quantile within block $j$. Running conditionally on $G = j$ yields $\theta_j - \theta_j^{\mathrm{loc}} \approx \theta_j s_j$, where
\begin{equation}
s_j \;=\; \frac{\E\bigl[Y \one\{Y > c_j\} \bigm| G = j\bigr]}{\E[Y \mid G = j]} ,
\qquad
\frac{\theta - \theta^{\mathrm{loc}}}{\theta}
\;\approx\; \frac{\sum_j \omega_j \theta_j s_j}{\sum_j \omega_j \theta_j} .
\label{eq:localshare}
\end{equation}
The global volume share is replaced by an effect-weighted average of \emph{within-block} volume shares. Because lower blocks have finite endpoints, their units cannot differ excessively, keeping $s_j$ small. The top block, a half-line, still contains the whole tail of the population. Blocking does not remove the problem, but confines it to one identifiable cell ($s_J$) where its size can be computed and reported, rather than spreading it silently across the global estimate. 

\section{Results}
\label{sec:numerics}

\paragraph{Design.} Pre-experiment scale $x$ is lognormal, truncated below at its $20$th percentile to represent enrollment filters, normalized to $\E[x] = 1$, and calibrated so that the top one percent of units hold $40\%$ of the total \citep{Hejazinia2019}. The outcome satisfies $\E[Y(0) \mid x] = x$ with variance $v(x) = 0.05\,x + 0.5\,x^{2}$ and is drawn as a gamma: the quadratic term is constant-coefficient-of-variation behaviour, and the linear term is counting noise. Blocks are $J = 20$ equal-probability cells. We take $N = 20{,}000$ and $2000$ replications, comparing six procedures under three structures: constant per-unit effects, a proportional lift in the top two blocks, and opposite signs (scaled so $\theta=0$). Table~\ref{tab:sim} reports bias, spread, and coverage.

\begin{table}[ht]
\centering\footnotesize
\caption{Bias, spread and coverage (nominal 95$\%$) over $2000$ replications. Design-weighted estimators are
unbiased throughout; the two that are not are the two that alter the estimand.}
\label{tab:sim}
\begin{tabular}{lrrrrrrrrr}
\toprule
Estimator & \multicolumn{3}{c}{constant, $\theta = 0.050$} & \multicolumn{3}{c}{tail-only, $\theta = 0.077$} & \multicolumn{3}{c}{opposite signs, $\theta = 0.000$} \\
\cmidrule(lr){2-4}\cmidrule(lr){5-7}\cmidrule(lr){8-10}
  & Bias & SD & Cov. & Bias & SD & Cov. & Bias & SD & Cov. \\
\midrule
Complete randomization & -0.0002 & 0.154 & 0.95 & +0.0041 & 0.186 & 0.96 & -0.0035 & 0.152 & 0.96 \\
\quad + global winsorization & -0.0004 & 0.029 & 0.95 & -0.0430 & 0.031 & 0.68 & +0.0124 & 0.028 & 0.93 \\
Blocked & -0.0028 & 0.151 & 0.95 & +0.0002 & 0.175 & 0.96 & -0.0001 & 0.146 & 0.96 \\
\quad + within-block winsorization & -0.0004 & 0.059 & 0.94 & -0.0122 & 0.060 & 0.95 & +0.0021 & 0.057 & 0.96 \\
\quad + within-block adjustment & -0.0007 & 0.076 & 0.89 & +0.0008 & 0.082 & 0.89 & -0.0012 & 0.073 & 0.90 \\
\quad + inverse-variance weights & +0.0000 & 0.001 & 0.95 & -0.0773 & 0.001 & 0.00 & +0.0039 & 0.001 & 0.03 \\
\bottomrule
\end{tabular}

\end{table}

\paragraph{How much precision is on the table.} Evaluating \eqref{eq:ratio} exactly for this population gives $1 - R^2_G = 0.958$, $0.935$, $0.860$ and $0.722$ at $J = 10$, $20$, $100$ and $1000$. At the granularity in common use, blocking removes about six percent of the variance, aligning with the modest 5--10\% improvements reported empirically. Table~\ref{tab:kappa} isolates $\kappa_J$, locating the difficulty there. At baseline concentration, twenty cells retain only a tenth of the covariate's explanatory power. Varying the coefficient on $x^2$ in $v(x)$ from $0.05$ to $1.00$ moves the ceiling $1 - R^2_x$ heavily, but moves $1 - R^2_G$ at $J = 20$ only marginally (from $0.906$ to $0.951$). When a partition retains a tenth of what the covariate knows, how much the covariate knows is close to irrelevant. The modest precision gain reported here is a consequence of granularity under concentration, not outcome noise.

\begin{table}[ht]
\centering\footnotesize
\caption{The discretization factor $\kappa_J$ of \eqref{eq:decomp}: the share of the covariate's
explanatory power that $J$ equal-count cells retain, by how concentrated the population is.}
\label{tab:kappa}
\begin{tabular}{lrrrrr}
\toprule
Top 1\% hold & $J=10$ & $J=20$ & $J=100$ & $J=1,000$ & $J=10,000$ \\
\midrule
20\% & 0.313 & 0.415 & 0.625 & 0.829 & 0.932 \\
40\% & 0.063 & 0.098 & 0.211 & 0.419 & 0.624 \\
60\% & 0.007 & 0.013 & 0.038 & 0.115 & 0.244 \\
\bottomrule
\end{tabular}

\end{table}

\paragraph{Winsorization and Weighting.} Global winsorization removes $56\%$ of the tail effect and covers at $0.68$. Winsorizing within block at the same nominal rate removes only $16\%$, restoring coverage to $0.95$ while still cutting standard deviation from $0.175$ to $0.060$. Meanwhile, inverse-variance weighting behaves as \eqref{eq:ivbias} requires: it drastically alters the estimand under tail effects, carrying $-99.9\%$ bias and placing merely one part in ten thousand of its weight on the top two blocks, despite reporting a tiny standard error. 

\paragraph{The block profile and qualifications.} Where the aggregate is uninformative, the profile is not. Under opposite signs, the top block shows a mean estimated $\theta_j$ of $-0.382$ against a true $-0.384$, while lower blocks show $+0.082$ against $+0.081$; the profile correctly reports a change that helps small units and harms large ones. Two qualifications attach to these results: within-block adjustment halves standard deviation but covers at $0.89$ (leverage-corrected variance recovers only about a third of the shortfall), and analysing blocked data as complete leaves the point estimate unchanged while overstating standard error by 6\% and lifting coverage to $0.98$. 

\section{Discussion and Future Research}
\label{sec:discussion}

The three benefits do not agree about how fine blocks should be. Precision and homogeneity improve as $J$ grows. The constraint comes from the other direction, and it is not a constraint on the design. Equation~\eqref{eq:variance} needs a within-arm sample variance in every block, requiring at least two units per arm per block, and is undefined at the matched-pair limit; within-block adjustment needs enough units to fit a slope. Assignment can be blocked as finely as the design permits, but the analysis cannot follow it down, and \eqref{eq:variance} is valid only for the partition the analysis actually uses. The floor on useful block size is therefore set by what can be estimated inside a block, not by what can be randomized inside one. This is the boundary \citet{bai2026efficiency} approach from the other side: at the fine limit the design supplies, unaided, what adjustment would otherwise provide.

Two downsides deserve stating plainly. Blocking must be committed to before randomization and can use only what is known then, whereas post-stratification may use any pre-treatment variable and be chosen after the fact. Furthermore, the benefits above are all conditional on a blocking variable that is predictive of both the outcome for \eqref{eq:ratio} and the effect for \eqref{eq:ivbias}. What recommends measures of pre-experiment scale in skewed populations is that they are often plausible candidates for both.

Two directions follow. The first is to stop allocating blocks uniformly. Concentrating blocks in the tail and leaving the body coarse should result in most of the available reduction at a fraction of the block count. The second follows from the constraint above: a variance estimator that credits fine assignment blocking while forming the point estimate on a coarser partition would let the design run ahead of the analysis, which is currently what prevents the gain from being reported even when it is real.

{\setlength{\bibsep}{2pt}
\bibliographystyle{abbrvnat}
\bibliography{refs}}

\end{document}